\documentclass[a4paper]{jpconf}

\usepackage{graphicx}
\usepackage{amsfonts}
\usepackage{amsmath}
\usepackage{amssymb}
\usepackage{mathrsfs}
\usepackage[english]{babel}
\usepackage[latin1]{inputenc}
\usepackage[T1]{fontenc}
\usepackage{epstopdf}
\usepackage{booktabs}

\newcommand{\dvfn}{\delta \! \varphi_{\!_{\perp}}}

\begin{document}

\title{Modif\mbox{}ied gravity, the Cascading DGP model and its critical tension}

\author{Fulvio Sbis\`a$^{1,2}$}

\address{$^1$ Institute of Cosmology and Gravitation, University of Portsmouth, UK}
\address{$^2$ Dipartimento di Fisica dell'Universit\`a di Milano and INFN, Sezione di Milano, Via Celoria 16, I-20133 Milano, Italy}

\ead{fulvio.sbisa@gmail.com}

\begin{abstract}
We investigate the presence of instabilities in the Cascading DGP model. We start by discussing the problem of the cosmological late time acceleration, and we introduce the modified gravity approach. We then focus on brane induced gravity models and in particular on the Cascading DGP model. We consider configurations of the latter model where the source term is given simply by vacuum energy (pure tension), and we study perturbations at first order around these configurations. We perform a four-dimensional scalar-vector-tensor decomposition of the perturbations, and show that, regarding the scalar sector, the dynamics in a suitable limit can be described by a master equation. This master equation contains an energy scale (critical tension) which is related in a non-trivial way to the parameters of the model. We give a geometrical interpretation of why this scale emerges, and explain its relevance for the presence of ghost instabilities in the theory. We comment on the difference between our result and the one present in the literature, and stress its importance regarding the phenomenological viability of the model. We finally provide a numerical check which confirms the validity of our analysis.
\end{abstract}

\section{Introduction}
More than 15 years ago, cosmologists had a satisfying picture of how the universe worked, based on three main hypotheses: it was assumed that the universe is homogeneous and isotropic on large scales, that energy and momentum exist in the form of ``ordinary'' matter and radiation plus Cold Dark Matter (CDM), and that gravity is described by Einstein's General Relativity (GR). This Standard Cosmological model was successful in explaining several aspects of our universe, and predicted that the expansion of the universe necessarily has to decelerate. In fact, by the second Friedmann equation, the scale factor of the Universe $a(t)$ obeys $\ddot{a}/a \propto - (\rho + 3 p)$ (where $\rho$ is the energy density, $p$ is the pressure and a dot indicates a time derivative), and the right hand side of the previous relation is negative for ordinary matter, radiation and CDM. The discovery of the present acceleration of the expansion \cite{Riess98, Perlmutter98}, successively confirmed by a variety of observations, forces us to abandon some of the assumptions mentioned above. We could relax the assumption of homogeneity, which differently from isotropy is not well-tested, and consider Lema\^{i}tre-Tolman-Bondi models where the Earth is placed inside a large void. Another possibility is that the backreaction of the deviations from homogeneity and isotropy on the evolution of the scale factor is not negligible, and neglecting this effect may cause a misinterpretation of the observational data. If instead we enforce homogeneity, we could postulate that the universe is filled by a source (which we don't observe in the lab or in the solar system) which has negative pressure: the simplest option is the Cosmological Constant ($\Lambda$), but we may also consider new dynamical degrees of freedom (the \emph{dark energy} paradigm). Finally, it may be that the cosmological observations are simply signalling that General Relativity is not the correct description of gravity at very large scales (the \emph{modified gravity} approach). See \cite{EllisNicolaiDurrerMaartens, DurrerMaartens} and references therein for a more detailed discussion of these approaches.

The most straightforward solution of this problem seems to allow for a non-zero cosmological constant (\emph{$\Lambda$CDM models}). However, fitting the observational data gives a value for $\Lambda$ which is really puzzling. If $\Lambda$ is considered as a new scale for gravity, then the universe is very fine tuned (coincidence problem), while if it is considered as the semi-classical manifestation of vacuum energy then the mismatch with the theoretical prediction is dramatic (``old'' cosmological constant problem) \cite{WeinbergCC}. Therefore, it seems worthwhile to pursue alternative approaches, and we choose to investigate the modified gravity approach.

\section{Modif\mbox{}ied gravity and Cascading DGP}

Modifying gravity is however potentially dangerous, since quite in general it introduces extra degrees of freedom compared to GR. On the one hand, these extra degrees of freedom have to be screened on terrestrial and astrophysical scales, otherwise they would be detected via fifth-force experiments and precise tests of gravity in the solar system. On the other hand, when studying the perturbative stability of certain solutions of the equations of motion, quite often some perturbation modes turn out to have negative kinetic energy and therefore lead to ghost instabilities. In fact, the presence of fields with negative kinetic energy (ghosts) renders the Hamiltonian unbounded from below, and causes the system to be unstable with respect to the simultaneous excitation of ghost and non-ghost fields (see \cite{Sbisa:2014pzo} and references therein). Several modified gravity theories have been proposed so far, including $f(R)$ gravity, massive gravity, and braneworlds (see \cite{ModifiedGravityAndCosmologyHugeReview} for a review). In particular, braneworld models are appealing from a fundamental point of view, since the presence of extra dimensions and branes is crucial in string theory. In these theories, the spacetime has more than four dimensions and matter and radiation (as well as the strong, weak and electro-magnetic interactions) are confined on structures (branes) whose dimensionality is lower than the dimensionality of the ambient spacetime (bulk). Only gravity can propagate in the extra dimensions, which is consistent with the fact that in string theory gravity is described by closed strings. The codimension of a brane is defined as the difference between the dimension of the bulk and the dimension of the brane, and the embedding functions are those functions which indicate the position in the bulk of the points belonging to the brane. In general the confinement is not sharp, so matter, radiation and the gauge fields are distributed around the brane within a characteristic distance, which is determined by the properties of the system and of the confinement mechanism. This characteristic length scale is called the thickness of the brane, while the details of the distribution constitute the internal structure of the brane.

Concerning the late time acceleration problem, a promising idea is that of brane induced gravity, whose first and best-known realization is the DGP model \cite{DGP00}. Braneworld theories with induced gravity are characterized by the inclusion, in the part of the action which describes the dynamics of the brane, of an Einstein-Hilbert term built from the metric induced on the brane. This term can be introduced at classical level purely on phenomenological grounds, but can be also understood as a contribution coming from loop corrections in the low energy effective action of a quantum description where matter is confined on the brane \cite{DGP00}. The DGP model has the intriguing property of admitting self-accelerating cosmological solutions, which open the possibility of explaining the cosmic acceleration by geometrical means. On the other hand, it has been shown that these solutions have a perturbative ghost instability, and that the DGP model fits the cosmological data significantly worse than $\Lambda$CDM. See \cite{Sbisa:2014dwa} and references therein for a discussion on the DGP model and its problems. A natural idea to go beyond the DGP model is to increase the codimension, while still having infinite extra dimensions and brane induced gravity. Such models offer also the possibility of addressing the Cosmological Constant problem, since they evade Weinberg's no-go theorem \cite{ArkaniHamed:2002fu, Dvali:2002pe} and may realize the degravitation mechanism \cite{Dvali:2007kt}. However, it is not clear if increasing the codimension helps with the ghost problem, since there are contradicting results \cite{Dubovsky:2002jm, Kolanovic:2003am, Berkhahn:2012wg}. Moreover, the gravitational field on a brane of codimension higher than one diverges when the thickness of the brane tends to zero \cite{Cline:2003ak, Vinet:2004bk, Bostock:2003cv}, and it is necessary to take explicitly into account its internal structure.

\subsection{The Cascading DGP model}

Both of the aforementioned problems were claimed to be solved in a fairly recent class of models, the Cascading DGP \cite{deRham:2007xp}. In these models there is a recursive embedding of branes into branes of increasing dimensionality, each equipped with an appropriate induced gravity term. In the minimal set-up, a four-dimensional (4D) brane (our universe) is embedded inside a 5D brane which in turn is embedded in the 6D bulk. This model has three parameters, the masses $M_6$, $M_5$ and $M_4$, where $M_6^4$ controls the strength of the bulk action, $M_5^3$ controls the strength of the induced gravity term on the 5D brane and $M_4^2$ controls the strength of the induced gravity term on the 4D brane. The requirement that the model reproduces Einstein gravity on small scales fixes $M_4$ to be equal to the Planck mass, so this model has in truth two free parameters, the mass scales $m_6 = M_6^4/M_5^3$ and $m_5 = M_5^3/M_4^2$. They respectively control the relative strength between the bulk action and the induced gravity term on the 5D brane ($m_6$), and between the induced gravity term on the 5D brane and the induced gravity term on the 4D brane ($m_5$). It was shown that, moving from large distances to small distances, weak gravity ``cascades'' $6D \to 5D \to 4D$ when $m_6  \ll m_5$ while there is a direct transition  $6D \to 4D$ when $m_6 \gg m_5$. Moreover, it was claimed that the presence of the codimension-1 brane with induced gravity renders the gravitational field finite on the codimension-2 brane even when the thickness of the latter tends to zero \cite{deRham:2007xp, deRham:2007rw}. A very important class of configurations are those that correspond to a source term given simply by vacuum energy ($\bar{\lambda}$) on the 4D brane (pure tension solutions). Importantly, it was claimed that there exists a critical tension $\bar{\lambda}_c$ such that a pure tension configuration is free of ghost instabilities (at first order in perturbations) if $\bar{\lambda} > \bar{\lambda}_c$ while it has ghosts if $\bar{\lambda} < \bar{\lambda}_c$ \cite{deRham:2007xp,deRham:2010rw}.

\section{The nested brane realization of the Cascading DGP model}
\label{regularization choice}

To probe the dynamics of the internal structure of a brane, we need to excite it with amounts of energy roughly of the order of the inverse of the thickness (in ``natural units'' $\hbar = c = 1$). If we are interested only in what happens outside of the brane, and want to focus on energy scales lower than the inverse of the thickness, it is usual to consider a ``thin limit description'' in which the thickness of the brane is sent to zero while keeping fixed the amount of energy and momentum on the brane. In this case the brane is said to be ``thin''. While this is extremely useful for codimension-1 branes, it was proved that the thin limit of branes of codimension higher than one is not well-defined \cite{GerochTraschen}. This result is very likely to be true also for more elaborate constructions such as in the Cascading case. Therefore, to obtain a well-defined formulation of the Cascading DGP model, it is necessary to give some information on the internal structure of the branes.

If we assume that there is a hierarchy between the thickness of the branes, namely that the codimension-1 brane is much thinner than the codimension-2 brane, we can describe the system as if a ``ribbon'' codimension-2 brane was present inside a thin codimension-1 brane. Furthermore, it can be shown that the thin limit of the ribbon brane (inside the already thin codimension-1 brane) is well defined \cite{Sbisa':2014uza}. It follows that, with this assumption about the hierarchy of the thicknesses, we can indeed work with thin branes and forget the internal structures. We refer to these set-ups as the \emph{nested branes realization} of the Cascading DGP.

\subsection{Perturbations around pure tension solutions}

To investigate the phenomenon which gives rise to the critical tension, we study perturbations at first order around solutions where pure tension is placed on the codimension-2 brane. These background solutions are most naturally expressed in a bulk-based approach, where the bulk metric is flat and the codimension-1 embedding has a cusp at the codimension-2 brane \cite{Sbisa':2014uza}. The complete space time is the product of a 4D Minkowski space and a 2D Riemannian cone, whose deficit angle is proportional to the tension. Therefore there exists a maximum tension $\bar{\lambda}_M = 2 \pi M_{6}^{4}$ which corresponds to a deficit angle of $2 \pi$ (when the cone becomes degenerate).

To study perturbations, we leave both the bulk metric and the codimension-1 embedding free to fluctuate. To deal with the issue of gauge invariance, we consider a 4D scalar-vector-tensor decomposition, and work with gauge-invariant variables \cite{Mukohyama:2000ui}. In particular, in the scalar sector it is possible to express the equations in terms of two master variables, the trace part $\pi$ of the bulk metric perturbations, and the normal component of the codimension-1 bending $\dvfn$ \cite{Sbisa':2014uza} (we call \emph{bending modes} the perturbation of the embedding functions). Notably, if we focus on the behaviour of the fields near the codimension-2 brane, it is possible to eliminate $\dvfn$ and obtain a master equation for $\pi$ alone. This equation however contains the derivative of $\pi$ normally to the codimension-1 brane, so to know the behaviour of $\pi$ on the codimension-2 brane it is still necessary to solve the full 6D problem. This difficulty can be overcome by considering a ``4D limit'', which gives the following (local) equation on the codimension-2 brane \cite{Sbisa:2014vva}
\begin{equation}
3 M_4^2 \, \bigg[ \, 1 - \frac{3}{2} \, \frac{m_5}{m_6} \, \tan \bigg( \frac{\bar{\lambda}}{4 M_6^4} \bigg) \bigg] \,\, \Box \, \pi = \mathcal{T}
\end{equation}
where $\Box$ indicates the four-dimensional D'Alembert operator and $\mathcal{T}$ is the trace of the energy-momentum tensor. This equation indicates that $\pi$ is an effective 4D ghost if $0 < \bar{\lambda} < \bar{\lambda}_c$ while it is a healthy field if $\bar{\lambda}_c < \bar{\lambda} < \bar{\lambda}_M$, where the \emph{critical tension} reads $\bar{\lambda}_c \equiv 4 M_6^4 \, \arctan \big( 2 m_6/3 m_5 \big)$.

\section{Geometrical interpretation and ghost-free regions in parameters space}

Having identified the critical tension, we can study the coupled dynamics of the fields $\pi$ and $\dvfn$ to interpret geometrically its existence. Considering the 4D limit mentioned above, it can be shown that the trace of the energy-momentum tensor $\mathcal{T}$ excites $\pi$ via two separate channels. It does so directly, because of the 4D induced gravity term, and indirectly via the bending of the codimension-1 brane, because of the 5D induced gravity term. Crucially, the first channel excites $\pi$ in a ghostly and $\bar{\lambda}$-independent way, while the second channel excites $\pi$ in a healthy and $\bar{\lambda}$-dependent way \cite{Sbisa:2014vva}. The existence of the critical tension is due to the competition between these two channels, and the fact that the field $\pi$ is a ghost or not is decided by the first channel being more or less efficient than the second channel. Note that the existence of the second channel is entirely due to the higher dimensional structure of the theory, and in particular to the presence of the induced gravity term on the codimension-1 brane.

Our result for the critical tension is at odds with the findings of \cite{deRham:2007xp,deRham:2010rw}, which found the value $\bar{\lambda}_{c}^{dRKT} = 8 \, m_6^2 \, M_4^2/3$. These two results coincide when $m_6 \ll m_5$ but differ significantly when $m_6 \gtrsim m_5$ and are dramatically different when $m_6 \gg m_5$. To see why this difference is crucial, note that the value we found for the critical tension is always smaller than the maximum tension $\bar{\lambda}_{M}$, so for every value of $m_5$ and $m_6$ we can find a range of values for the background tension such that $\pi$ is a healthy field. However, $\bar{\lambda}_{c}^{dRKT}$ is smaller than $\bar{\lambda}_{M}$ only when $m_6 \lesssim m_5$, so the results of \cite{deRham:2007xp,deRham:2010rw} imply that half of the phase space of the theory is plagued by ghosts, and so is phenomenologically ruled out. It is therefore very important to establish why two different results are obtained, and which of the two is correct.

\subsection{Numerical check}

A priori we may wonder if, referring to the discussion in section \ref{regularization choice}, the hypotheses used in \cite{deRham:2007xp,deRham:2010rw} about the internal structure of the branes are different from the one we use, and therefore in truth we are considering different models. However this is not the case, since our analysis can be mapped into that of \cite{deRham:2010rw} by a coordinate transformation \cite{Sbisa:2014vva}. In fact it can be shown that the difference lies in how the junction conditions at the codimension-2 brane are derived, which is linked to which hypotheses are made on the behaviour of the fields near the codimension-2 brane in the thin limit. Roughly speaking, we assume that the embedding functions remain continuous in the thin limit (``the codimension-1 brane does not break''), while the result of \cite{deRham:2010rw} is reproduced assuming that the normal component of the bending remains continuous. These two conditions cannot be both satisfied at the same time, since in the background solutions the normal vector becomes discontinuous in the thin limit \cite{Sbisa:2014vva}. To decide which of the two results is correct, we consider a case, the pure tension perturbation case, where the internal structure of the codimension-2 brane is exactly solvable. In this case it is possible to derive the codimension-2 junction conditions by performing numerically the integration of the codimension-1 junction conditions across the codimension-2 brane (pillbox integration). Since the configuration of the fields is known explicitly also inside the codimension-2 brane, it is not necessary to make any hypothesis on the behaviour of the fields to do that.

\begin{figure}[htp!]
\begin{center}
\includegraphics{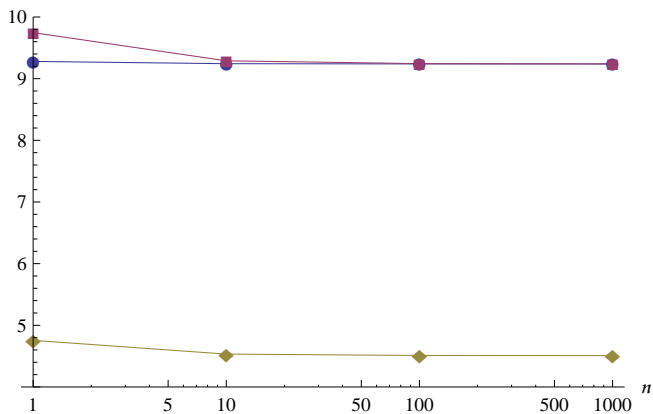}
\caption[Numerical results of the pillbox integration]{Plot of the numerical results of the pillbox integration}
\label{PillboxIntegrationfigure}
\end{center}
\end{figure}

The idea is to perform the numerical pillbox integration for a sequence of configurations with different thickness, and study the asymptotic behaviour of the outcome when the thickness tends to zero. If our analysis or the analysis of \cite{deRham:2007xp,deRham:2010rw} is correct, then it has to agree with the numerical outcome in the thin limit. The outcome of the numerical integrations are plotted in figure \ref{PillboxIntegrationfigure} (the parameter $n$ is inversely proportional to the thickness). It is evident that the points corresponding to our codimension-2 junction conditions (squares) converge to the points corresponding to the numerical integration (circles), while the points corresponding to the codimension-2 junction conditions which reproduce the result of \cite{deRham:2010rw} (diamonds) are significantly distant from the former ones. 

\section{Conclusions}

The numerical check above puts on firm footing our analysis of the nested branes realization of the 6D Cascading DGP, and strongly supports our claim that the correct value of the critical tension is $\bar{\lambda}_c \equiv 4 M_6^4 \, \arctan \big( 2 m_6/3 m_5 \big)$. In particular, it strongly suggests that also models where gravity displays a direct transition 6D $\to$ 4D ($m_6 > m_5$) are phenomenologically viable. We conclude that braneworld models with infinite volume extra dimensions and induced gravity might be a powerful tool to tackle the late time acceleration and the Cosmological Constant problems. In particular, the Cascading DGP seems a promising candidate to overcome the problems of the DGP models whilst preserving its good features. On the other hand, the singular structure of the geometry is very subtle at the codimension-2 brane, thus indirectly confirming the belief that the singular structure of branes of codimension higher than one is in general more complex than that of codimension-1 branes. 

Regarding future directions of research, a lot of work is still to be done, including studying the existence of the ghost at full non-linear level by performing a Hamiltonian analysis. More in general, it is important to derive the codimension-2 junction conditions at non-perturbative level and to derive cosmological solutions, as well as verifying if the model passes the high-precision solar system tests.

\ack This talk is based on work done in collaboration with Kazuya Koyama.

\section*{References}

\providecommand{\newblock}{}


\begin{thebibliography}{10}
\expandafter\ifx\csname url\endcsname\relax
  \def\url#1{{\tt #1}}\fi
\expandafter\ifx\csname urlprefix\endcsname\relax\def\urlprefix{URL }\fi
\providecommand{\eprint}[2][]{\url{#2}}

\bibitem{Riess98}
Riess A~G {\em et~al.\/} (Supernova Search Team) 1998 {\em Astron.~J.\/} {\bf 116} 1009--38  (\textit{Preprint} \eprint{astro-ph/9805201})

\bibitem{Perlmutter98}
Perlmutter S {\em et~al.\/} (Supernova Cosmology Project) 1999 {\em Astrophys.~J.\/} {\bf 517} 565--86  (\textit{Preprint} \eprint{astro-ph/9812133})

\bibitem{EllisNicolaiDurrerMaartens}
Ellis G, Nicolai H, Durrer R and Maartens R 2008 {\em Gen.~Rel.~Grav.\/} {\bf 40} 219--20

\bibitem{DurrerMaartens}
Durrer R and Maartens R 2008 {\em Dark Energy: Observational \& Theoretical Approaches\/} ed.~P Ruiz-Lapuente (Cambridge UP, 2010) 48--91 (\textit{Preprint} \eprint{0811.4132})

\bibitem{WeinbergCC}
Weinberg S 1989 {\em Rev.~Mod.~Phys.\/} {\bf 61} 1--22

\bibitem{Sbisa:2014pzo}
Sbisa' F 2015 {\em Eur.~J.~Phys.\/} {\bf 36} 015009 (\textit{Preprint} \eprint{1406.4550})

\bibitem{ModifiedGravityAndCosmologyHugeReview}
Clifton T, Ferreira P~G, Padilla A and Skordis C 2012 {\em Phys.~Rept.\/} {\bf 513} 1--189  (\textit{Preprint} \eprint{1106.2476})

\bibitem{DGP00}
Dvali G, Gabadadze G and Porrati M 2000 {\em Phys.~Lett.~B\/} {\bf 485} 208--214  (\textit{Preprint} \eprint{hep-th/0005016})

\bibitem{Sbisa:2014dwa}
Sbisa' F 2014 Modified theories of gravity \textit{Preprint} \eprint{1406.3384}

\bibitem{ArkaniHamed:2002fu}
Arkani-Hamed N, Dimopoulos S, Dvali G and Gabadadze G 2002 Nonlocal modification of gravity and the cosmological constant problem \textit{Preprint} \eprint{hep-th/0209227}

\bibitem{Dvali:2002pe}
Dvali G, Gabadadze G and Shifman M 2003 {\em Phys.~Rev.~D\/} {\bf 67} 044020  (\textit{Preprint} \eprint{hep-th/0202174})

\bibitem{Dvali:2007kt}
Dvali G, Hofmann S and Khoury J 2007 {\em Phys.~Rev.~D\/} {\bf 76} 084006  (\textit{Preprint} \eprint{hep-th/0703027})

\bibitem{Dubovsky:2002jm}
Dubovsky S and Rubakov V 2003 {\em Phys.~Rev.~D\/} {\bf 67} 104014  (\textit{Preprint} \eprint{hep-th/0212222})

\bibitem{Kolanovic:2003am}
Kolanovic M, Porrati M and Rombouts J~W 2003 {\em Phys.~Rev.~D\/} {\bf 68} 064018  (\textit{Preprint} \eprint{hep-th/0304148})

\bibitem{Berkhahn:2012wg}
Berkhahn F, Hofmann S and Niedermann F 2012 {\em Phys.~Rev.~D\/} {\bf 86} 124022  (\textit{Preprint} \eprint{1205.6801})

\bibitem{Cline:2003ak}
Cline J~M, Descheneau J, Giovannini M and Vinet J 2003 {\em J.~High Energy Phys.\/} JHEP06(2003)048 (\textit{Preprint} \eprint{hep-th/0304147})

\bibitem{Vinet:2004bk}
Vinet J and Cline J~M 2004 {\em Phys.~Rev.~D\/} {\bf 70} 083514  (\textit{Preprint} \eprint{hep-th/0406141})

\bibitem{Bostock:2003cv}
Bostock P, Gregory R, Navarro I and Santiago J 2004 {\em Phys.~Rev.~Lett.\/} {\bf 92} 221601  (\textit{Preprint} \eprint{hep-th/0311074})

\bibitem{deRham:2007xp}
de~Rham C, Dvali G, Hofmann S, Khoury J, Pujolas O {\em et~al.\/} 2008 {\em Phys.~Rev.~Lett.\/} {\bf 100} 251603  (\textit{Preprint} \eprint{0711.2072})

\bibitem{deRham:2007rw}
de~Rham C, Hofmann S, Khoury J and Tolley A~J 2008 {\em J.~Cosmol.~Astropart.~Phys.\/} JCAP02(2008)011  (\textit{Preprint} \eprint{0712.2821})

\bibitem{deRham:2010rw}
de~Rham C, Khoury J and Tolley A~J 2010 {\em Phys.~Rev.~D\/} {\bf 81} 124027  (\textit{Preprint} \eprint{1002.1075})

\bibitem{GerochTraschen}
Geroch R and Traschen J 1987 {\em Phys.~Rev.~D\/} {\bf 36} 1017

\bibitem{Sbisa':2014uza}
Sbisa' F and Koyama K 2014 {\em J.~Cosmol.~Astropart.~Phys.\/}  JCAP06(2014)029  (\textit{Preprint} \eprint{1404.0712})

\bibitem{Mukohyama:2000ui}
Mukohyama S 2000 {\em Phys.~Rev.~D\/} {\bf 62} 084015

\bibitem{Sbisa:2014vva}
Sbisa' F and Koyama K 2014 {\em J.~Cosmol.~Astropart.~Phys.\/} JCAP09(2014)038 (\textit{Preprint} \eprint{1405.7617})

\end{thebibliography}
\end{document}